\documentclass[prb,twocolumn]{revtex4-1}
\usepackage[latin9]{inputenc}
\usepackage{amstext}
\usepackage{graphicx}
\usepackage{xcolor}
\usepackage{amsfonts}
\usepackage{amsmath, mathtools}
\numberwithin{equation}{section}

\usepackage[colorlinks=true,citecolor=blue,
linkcolor=blue,urlcolor=blue]{hyperref}


\begin{document}

\title{The bound-state solutions of the one-dimensional hydrogen atom }

\author{Rufus Boyack and Frank Marsiglio}
\affiliation{Department of Physics and Theoretical Physics Institute, University
of Alberta, Edmonton, Alberta T6G 2E1, Canada}

\begin{abstract}
The one-dimensional hydrogen atom is an intriguing quantum mechanics
problem that exhibits several properties which have been continually
debated. In particular, there has been variance as to whether or not
even-parity solutions exist, and specifically whether or not the ground
state is an even-parity state with infinite negative energy. We study a ``regularized"
version of this system, where the potential is a constant in the
vicinity of the origin, and we discuss the even- and odd-parity solutions for this 
regularized one-dimensional hydrogen atom. We show how the even-parity states, with the
exception of the ground state, converge to the same functional form and become degenerate
for $x > 0$ with the odd-parity solutions as the cutoff approaches zero.
This differs with conclusions
derived from analysis of the singular (i.e., without regularization) one-dimensional 
Coulomb potential, where even-parity solutions are absent from the spectrum. 
\end{abstract}
\maketitle

\section{Introduction}

The one-dimensional (1D) hydrogen atom, described by the potential
$V\left(x\right)\sim1/\text{\ensuremath{\left|x\right|}}$, has long
been a quantum mechanics problem of theoretical interest, in part
due to a lack of consensus as to whether or not even-parity solutions
are admissible. There is a bevy of literature on this subject, and
for our purposes we highlight only several pertinent articles which
can be further investigated. The first study
of the odd-parity solutions of the 1D hydrogen atom was performed
in Ref.~[\onlinecite{Flugge_1952}],
where the usual hydrogen spectrum was derived. Following this initial study, a seminal
investigation by Loudon~\citep{Loudon_1959,Loudon_2016} used a regularization procedure to derive both
odd-parity and even-parity solutions; in particular,
Loudon found that the ground-state solution of the 1D hydrogen atom
is an even-parity function with infinite negative energy, whose square
modulus limits to a delta function as the regularization parameter goes to zero. Moreover, apart from his proposed
ground-state solution, Loudon found that all the other states are two-fold
degenerate, having either even or odd parity. There are two fundamental
questions related to Loudon's results which have caused disagreement
in the literature -- (i) are the solutions to the 1D hydrogen atom
degenerate; that is, are both the even-parity and odd-parity solutions
mathematically well-defined solutions obeying all necessary boundary
conditions? (ii) if even-parity solutions are permissible, is the
ground-state solution the same as that given by Loudon, which has
the peculiar property that its square modulus limits to a delta function?
Standard texts~\citep{LL_Book} in quantum mechanics argue that the bound-state
solutions of a 1D potential are non-degenerate; whether this
result remains valid for singular potentials requires a more elaborate analysis. 

In Ref.~[\onlinecite{Andrews_1966}],
Andrews further investigated Loudon's
ground-state solution, and there he showed that its scalar product
with any square-integrable function vanishes, upon which he concluded
that this solution is not observable and therefore presumably does not exist. In this mathematical analysis
it was important to take the limiting form of Loudon's ground-state
solution after performing the integration. This conclusion was similarly obtained by N$\acute{\text{u}}\tilde{\text{n}}$ez-Y$\acute{\text{e}}$pez
and Salas-Brito,~\citep{Yepez_1987}
who argued that the discrepancy with Ref.~[\onlinecite{Loudon_1959}]
is because Loudon computed the limiting probability distribution as opposed
to the limiting wave function. Their conclusion was that the ground-state solution therefore does not exist.
Haines and Roberts~\citep{Haines_1969} used the same regularized potential as Loudon and reached similar conclusions for this variant of the problem. 
However, for the singular Coulomb potential, they showed that only the odd-parity solutions are admissible. 
As a result, this ruled out any possible degeneracy
in the bound-state solutions. In addition, these authors also
claimed that there are a continuum of negative energy states. 

Andrews~\citep{Andrews_1976} later investigated the notion of degeneracy in singular potentials
and provided a more encompassing analysis than that usually given
in textbooks. Three types of potentials were defined by Andrews --
mildly singular (MS), singular (S), and extremely singular (XS). An
MS potential is one which is continuous, infinitely large at the origin,
but integrable: $\int_{0}^{L}\left|V\right|dx<\infty$. An S potential
is continuous, infinitely large at the origin, and nonintegrable near
the origin: $\int_{0}^{L}x\left|V\right|dx<\infty$ but $\int_{x}\left|V\left(y\right)\right|dy\rightarrow\infty$
as $x\rightarrow0$. An XS potential is continuous with $\int_{0}x\left|V\right|dx$
being divergent. Andrews showed that MS potentials behave in the same
manner as non-singular potentials -- their wave functions and the first
derivatives of the wave functions are continuous. However, he argued
that for class S potentials the potential acts as an impenetrable barrier and hence
there is no reason to try to match the wave functions at boundaries.
As a result, the even-parity bound state solutions, which Haines and
Roberts deemed inadmissible based on a rigorous continuity condition
derived from the Schr$\ddot{\text{o}}$dinger equation, were argued
by Andrews to be admissible; the reason being that whilst such solutions
add a delta function at the origin, it is claimed that this does not
affect any matrix elements of the Hamiltonian. In addition, 
Andrews argued that the continuum of negative energy states discussed by Haines
and Roberts is incorrect. A similar conclusion about the invalidity of Haines and Roberts' continuum solutions, and the absence of even-parity solutions, 
was reached by Gomes and Zimerman in Ref.~[\onlinecite{Gomes_1980}]; 
moreover, these latter authors asserted that the impenetrable boundary idea of Andrews is incorrect; see Refs.~[\onlinecite{Andrews_1981,Gomes_1981,Andrews_1988, Hammer_1988}].

More recently, Xianxi {\it et al.}~\citep{Xianxi_1997} revisited the
1D hydrogen atom, deriving a continuity condition for the derivative
of the wave function and showing that only odd-parity solutions therein
satisfy this condition. These authors also considered a similar argument
to that of Andrews~\citep{Andrews_1966} as to the reason for the
invalidity of Loudon's ground-state solution. Palma and Raff~\citep{Palma_2006}
argued that there is no degeneracy in the 1D hydrogen atom -- all
of the bound-state solutions have odd parity -- and moreover they
asserted that Loudon's ground-state wave function does not satisfy the
Schr$\ddot{\text{o}}$dinger equation. 
This resolves the disputes in the literature, providing definitive answers to the two
questions posed earlier in the introduction, and disproves the claim by Andrews
that the (singular) 1D hydrogen atom acts as an impenetrable
barrier. Further references on methods of solution to the 1D hydrogen
atom can be found in Ref.~[\onlinecite{Palma_2006}] and a detailed collection of additional references is in Ref.~[\onlinecite{Fischer_1995}].
Another collection of recent literature~\citep{Jaramillo_2009,Yepez_2011} has claimed
that there are neither even nor odd-parity eigenstates, the argument~\citep{Yepez_2014}
being that the odd-parity states would necessarily have vanishing
expectation values for both the position and momentum operators. However,
this is not a mathematical justification for discarding the odd-parity solutions. 

Other approaches to studying the one-dimensional Coulomb potential include the use of symmetry arguments,~\citep{Davtyan_1987,Ivetic_2018} the Laplace transform,~\citep{Gordeyev_1997}
and the theory of distributions.~\citep{Fischer_1995,Kurasov_1996,Calcada_2014,Calcada_2019,Golovaty_2019}
The $d$-dimensional hydrogen atom has been studied also,~\citep{Nieto_1979,Moss_1987} although the problems associated with the even-parity solutions in
the $d=1$ case were not fully addressed. 

The 1D hydrogen atom system has also garnered interest~\citep{Grimes_1976,YK_Book}
in its application to the explanation of why electrons ``float''
above the surface of liquid helium -- in one of the simplest models,
the bulk of the inert helium acts as an infinite barrier while in
the region outside the helium an electron experiences an attractive
force with an image charge created in the dielectric medium. The
resulting calculation of the mean distance an electron floats above
the liquid helium surface is in reasonable agreement with experiment.~\citep{Grimes_1976} 
In this 1D hydrogen-atom-like potential, due to the infinite barrier in half of the region of space, 
the wave function must necessarily vanish in this region, and thus 
the issue of whether there are even or odd-parity eigenstates is unimportant. 

In this paper we consider a well-defined
regularized 1D Coulomb potential and analyze
the behavior of the even- and odd-parity energy eigenstates as we remove the cutoff. We
see that in this limit the even-parity solutions develop a cusp at the origin. Hence, while
valid solutions of the regularized system, they do not tend to valid solutions of the system
with a singular Coulomb potential as the limiting forms do not satisfy the criterion of
having a continuous first derivative.\cite{Palma_2006}
We use a constant potential
near the origin up to some cutoff, followed by the Coulomb potential. One can then take this cutoff to be as small as one wishes.
While differences with the results for the ``bare'' Coulomb potential may well exist (as reviewed above) we argue that the results following
this procedure are the most physical ones, given that any hydrogen potential is usually generated by a central nucleus of finite extent.
Thus our results will be very similar to those of Loudon, with both even and odd-parity solutions existing, and an even ground state whose energy
becomes increasingly negative with decreasing cutoff. We do not contribute further to the non-regularized problem -- all results quoted here come primarily from
Andrews\cite{Andrews_1966,Andrews_1976} and Palma and Raff.\cite{Palma_2006} The essential point is that our regularized calculations always have the ground state
and the ensuing excited even states. The existence of such a ground state acts as a (penetrable) barrier for the remaining
eigenstates, and indeed, for any function constructed from them.\citep{Ibrahim_2018} 

The outline of the paper is as follows. In Sec.~\ref{sec:RegCoulomb} we study our particular regularized Coulomb potential.
The energy eigenvalue equations for both the even and odd-parity solutions are investigated in Sec.~\ref{sec:EnergyEigenvalues},
and, following this, in Sec.~\ref{sec:Delta0Limits} we provide analytical expressions for the energy eigenvalues in 
the limit that the regularized potential approaches the singular potential. 
The wave functions are investigated in Sec.~\ref{sec:Wavefunctions} and finally we conclude in Sec.~\ref{sec:Conclusions}.

\section{Regularized Coulomb potential formalism}
\label{sec:RegCoulomb}

In this section we consider a regularized one-dimensional hydrogen
atom potential given by
\begin{equation}
V\left(x\right) =
\begin{dcases}
       -V_{0}\frac{a_{0}\delta}{\left|x\right|}, &  \left|x\right|\geq\delta a_{0} \\
       -V_{0}, & \left|x\right|\leq\delta a_{0}.    
\end{dcases}
\label{eq:Potential}
\end{equation}
Here $a_{0}=\frac{4\pi\epsilon_{0}\hbar^{2}}{me^{2}}$ is the Bohr
radius and $\delta>0$ is a positive constant. The constant $V_{0}$ is defined by 
\begin{equation}
V_{0}=\frac{e^{2}}{4\pi\epsilon_{0}a_{0}\delta}.
\end{equation}
Loudon~\citep{Loudon_1959} studied a different regularization, but he provided some brief comments and calculations for this potential as well. Here we
provide a more in-depth analysis of the potential in Eq.~\eqref{eq:Potential}, and in particular
we follow the method of analysis of Ref.~[\onlinecite{Othman_2017}], which
highlighted how to solve the conventional hydrogen atom in a more
symmetric fashion when considering the boundary conditions near the
origin and near infinity. Since the potential is parity-invariant,
the solutions to the Schr$\ddot{\text{o}}$dinger equation have definite parity.
Let region I be defined by $\left|x\right|\leq\delta a_{0}$ whereas region
II is $\left|x\right|\geq a_{0}\delta$. The Schr$\ddot{\text{o}}$dinger
equation in region I is then 
\begin{equation}
-\frac{\hbar^{2}}{2m}\psi^{\prime\prime}\left(x\right)-V_{0}\psi\left(x\right)=E\psi\left(x\right).\label{eq:SE_I}
\end{equation}
In what follows we focus on the bound-state solutions only. Define the wave vector $q$, the energy $E$, and the coordinate $y$ by
\begin{equation}
q=\sqrt{\frac{2m}{\hbar^{2}}\left(E+V_{0}\right)},\ \ E=-\frac{\hbar^{2}}{2ma_{0}^{2}\beta^{2}}, \ \ y\ =\frac{x}{a_{0}\beta}.\label{eq:Vars}
\end{equation}
Note that 
\begin{equation}
q=\frac{1}{a_{0}}\sqrt{\left(\frac{2}{\delta}-\frac{1}{\beta^{2}}\right)}.
\end{equation}
Thus, Eq.~\eqref{eq:SE_I} now becomes 
\begin{equation}
\psi^{\prime\prime}\left(x\right)+q^{2}\psi\left(x\right)=0.
\end{equation}
The general solution to this differential equation is 
\begin{equation}
\psi\left(x\right)=A_{\text{I}}\cos\left(qx\right)+B_{\text{I}}\sin\left(qx\right).
\end{equation}
The even-parity solution has $B_{\text{I}}=0$ whereas the odd-parity
solution has $A_{\text{I}}=0$. In region II, the Schr$\ddot{\text{o}}$dinger
equation is 
\begin{equation}
-\frac{\hbar^{2}}{2m}\psi^{\prime\prime}\left(x\right)-\frac{e^{2}}{4\pi\epsilon_{0}\left|x\right|}\psi\left(x\right)=E\psi\left(x\right).\label{eq:SE_II}
\end{equation}
Rewriting Eq.~\eqref{eq:SE_II} in terms of the variables introduced
in Eq.~\eqref{eq:Vars} gives 
\begin{equation}
\psi^{\prime\prime}\left(y\right)+\frac{2\beta}{\left|y\right|}\psi\left(y\right)-\psi\left(y\right)=0.
\end{equation}
Consider the region $y\geq0$; let $\psi\left(y\right)=ye^{-y}f\left(y\right)$
and substitute this into the above differential equation to obtain
\begin{equation}
yf^{\prime\prime}+2\left(1-y\right)f^{\prime}-2\left(1-\beta\right)f=0.\label{eq:SEf}
\end{equation}
The Kummer equation is given by~\cite{AS_Book} 
\begin{equation}
yw^{\prime\prime}+\left(b-y\right)w^{\prime}-aw=0,
\end{equation}
and its solutions are~\cite{AS_Book} 
\begin{equation}
w\left(y\right)=cM\left(a,b;y\right)+dU\left(a,b;y\right).
\end{equation}
In Appendix~\ref{sec:App_Kummer} we provide a brief overview of the Kummer function $M$ and the Tricomi function $U$. 
The general solution to Eq.~\eqref{eq:SEf} is thus 
\begin{equation}
f\left(y\right)=A_{\text{II}}M\left(1-\beta,2;2y\right)+B_{\text{II}}U\left(1-\beta,2;2y\right).
\end{equation}
As $y\rightarrow\infty$, $M\left(a,b;y\right)\rightarrow\left[\Gamma\left(b\right)/\Gamma\left(a\right)\right]y^{a-b}e^{y}$, assuming that $a$ is not a negative integer; 
in the case that it is, the solution truncates to a power series. 
Proceeding under the assumption that $1-\beta$ is not a negative integer, which will turn out to be the case, 
then normalizability of the wave function requires $A_{\text{II}}=0$. 

\section{Energy eigenvalues}
\label{sec:EnergyEigenvalues}

\subsection{Even-parity solutions}

Consider the even-parity solution: $B_{\text{I}}=0$. Only the domain
$x\geq0$ then needs to be investigated, and the solution for negative
$x$ can be determined from $\psi\left(x\right)=\psi\left(-x\right)$.
The wave function is thus 
\begin{equation}
\psi\left(x\right) =
\begin{dcases}
 A_{\text{I}}\cos\left(qx\right), & x\leq\delta a_{0} \\
 B_{\text{II}}\frac{x}{a_{0}\beta}e^{-x/\left(a_{0}\beta\right)}U\left(1-\beta,2;\frac{2x}{a_{0}\beta}\right), & x\geq\delta a_{0}.
 \end{dcases}
\end{equation}
To determine the eigenvalue condition we match $\psi$ and $\psi^{\prime}$
at $x=\delta a_{0}$. In region I, $\psi^{\prime}/\psi$ at $x=\delta a_{0}$
is given by 
\begin{equation}
\frac{\psi_{\text{I}}^{\prime}}{\psi_{\text{I}}}=-q\frac{\sin\left(q\delta a_{0}\right)}{\cos\left(q\delta a_{0}\right)}.\label{eq:Ratio1}
\end{equation}
In the small $\delta$ limit this reduces to $-q^{2}\delta a_{0}=-{2}/{a_{0}}$,
which agrees with Eq.~(3.36) in Ref.~[\onlinecite{Loudon_1959}].  
To evaluate the derivatives of $\psi$ in region II, a useful identity is~\cite{AS_Book}
\begin{equation}
\frac{d}{dx}U\left(a,b;x\right)=-aU\left(a+1,b+1;x\right).
\end{equation}
Using this relation, in region II we find that $\psi^{\prime}/\psi$ at
$x=\delta a_{0}$ is given by 
\begin{equation}
\frac{\psi_{\text{II}}^{\prime}}{\psi_{\text{II}}}=\frac{1}{a_{0}\delta}\left[1-\frac{\delta}{\beta}-\frac{2\delta}{\beta}\left(1-\beta\right)\frac{U\left(2-\beta,3;\frac{2\delta}{\beta}\right)}{U\left(1-\beta,2;\frac{2\delta}{\beta}\right)}\right].\label{eq:Ratio2}
\end{equation}
Matching Eq.~\eqref{eq:Ratio1} and Eq.~\eqref{eq:Ratio2}, for a given
$\delta$, then determines the eigenvalue condition for $\beta$ and
thus $E$. The result is the following transcendental equation for
$\beta$:
\begin{equation}
-qa_{0}\delta\tan\left(qa_{0}\delta\right)=1-\frac{\delta}{\beta}-\left(1-\beta\right)\frac{2\delta}{\beta}\frac{U\left(2-\beta,3;\frac{2\delta}{\beta}\right)}{U\left(1-\beta,2;\frac{2\delta}{\beta}\right)},\label{eq:Energy1}
\end{equation}
where $q$ is defined in Eq.~\eqref{eq:Vars}. This can be rewritten
in a simplified manner by using the recurrence relations for the $U$ function. 
In particular, by using Eqs.~\eqref{eq:Urel1} and \eqref{eq:Urel2} in Appendix~\ref{sec:App_Kummer}, we find that 
\begin{align}
&\frac{2\delta}{\beta}U\left(2-\beta,3;\frac{2\delta}{\beta}\right) \nonumber\\
& =  U\left(1-\beta,2;\frac{2\delta}{\beta}\right)+\beta U\left(2-\beta,2;\frac{2\delta}{\beta}\right)\nonumber \\
 & =  \frac{1}{1-\beta}\left[U\left(1-\beta,2;\frac{2\delta}{\beta}\right)-\beta U\left(1-\beta,1;\frac{2\delta}{\beta}\right)\right].\label{eq:TricomiIdentity}
\end{align}
Inserting this identity into Eq.~\eqref{eq:Energy1} then simplifies
the even-parity eigenvalue equation to 
\begin{eqnarray}
qa_{0}\tan\left(qa_{0}\delta\right) & = & \frac{1}{\beta}\left[1-\frac{\beta^{2}}{\delta}\frac{U\left(1-\beta,1;\frac{2\delta}{\beta}\right)}{U\left(1-\beta,2;\frac{2\delta}{\beta}\right)}\right].\label{eq:EvenEnergy}
\end{eqnarray}

\subsection{Odd-parity solutions}

For the odd-parity solution: $A_{\text{I}}=0$. Only the domain $x\geq0$
then needs to be investigated, and the solution for negative $x$
can be determined from $\psi\left(x\right)=-\psi\left(-x\right)$.
The wave function is thus 
\begin{equation}
\psi\left(x\right) =
\begin{dcases}
B_{\text{I}}\sin\left(qx\right), & x\leq\delta a_{0}  \\
B_{\text{II}}\frac{x}{a_{0}\beta}e^{-x/\left(a_{0}\beta\right)}U\left(1-\beta,2;\frac{2x}{a_{0}\beta}\right), & x\geq\delta a_{0}.
\end{dcases}
\end{equation}
In region I, $\psi^{\prime}/\psi$ at $x=\delta a_{0}$ is given by
\begin{equation}
\frac{\psi_{\text{I}}^{\prime}}{\psi_{\text{I}}}=q\frac{\cos\left(q\delta a_{0}\right)}{\sin\left(q\delta a_{0}\right)}.
\end{equation}
Using the identity in Eq.~\eqref{eq:TricomiIdentity}, 
the odd-parity eigenvalue equation then becomes 
\begin{equation}
-qa_{0}\cot\left(qa_{0}\delta\right)=\frac{1}{\beta}\left[1-\frac{\beta^{2}}{\delta}\frac{U\left(1-\beta,1;\frac{2\delta}{\beta}\right)}{U\left(1-\beta,2;\frac{2\delta}{\beta}\right)}\right].\label{eq:OddEnergy}
\end{equation}

A full numerical solution of Eq.~\eqref{eq:EvenEnergy} and Eq.~\eqref{eq:OddEnergy} for the even and odd-parity cases, respectively, is given in the figures below. 
First, however, we explore some analytical results in the limit $\delta\rightarrow0$.

\section{Analytical results in the limit $\delta\rightarrow0$}
\label{sec:Delta0Limits}

\subsection{Even-parity solutions}

\begin{figure}
\centering
\includegraphics[width=10cm]{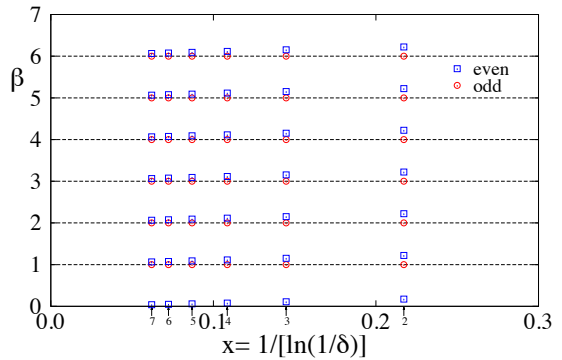}
\caption{The energy eigenvalues, $E_{n}$, for both even-parity (blue squares)
and odd-parity (red circles) eigenstates as a function of the cutoff,
$\delta a_{0}$, for the Coulomb potential. Motivated by Eq.~\eqref{eq:RegE},
we plot $\beta\equiv\left(-E_{0}/E_{n}\right)^{\frac{1}{2}}$ versus
$x\equiv1/\left[\ln\left(1/\delta\right)\right]$, where $E_{0}=-\frac{\hbar^{2}}{2ma_{0}^{2}}$.
An infinite spectrum of bound states arises, with all of the
excited-state energies trending towards the positive integers as $\delta\rightarrow0$.
On the bottom axis, small arrows accompanied with labels indicate
the actual value of $\delta$; the integer value indicates the negative
exponent of a power of 10 (e.g., the right-most set of points are
calculated for $\delta=10^{-2}$ while the left-most set of points
is for $\delta=10^{-7}$). The energy eigenvalues for the odd-parity
solutions converge to the integer values, even for ``large'' values
of $\delta$. However, the even-parity solutions are approaching
the integer values more slowly as $\delta\rightarrow0$. 
The even ground-state energy eigenvalue approaches $\beta\rightarrow0$ as $\delta\rightarrow0$. These trends will be examined
more closely in subsequent figures.}
\label{fig:Energy}
\end{figure}

\begin{figure}[t]
\centering
\includegraphics[width=10cm]{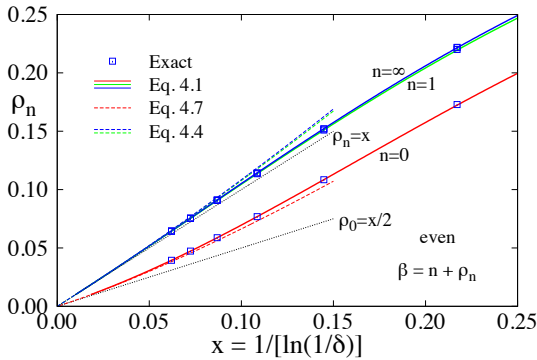}
\caption{The deviation of $\beta$ with respect to integer values, defined
as $\rho_{n}\equiv\beta-n$, versus $x\equiv1/\left[\ln\left(1/\delta\right)\right]$
for all the even-parity solutions. The exact values are given by the
squares -- these are the same results as shown in Fig.~\ref{fig:Energy}.
The solid curves are the solutions to Eq.~\eqref{eq:RegE}. 
The red curve is for the bound state $n=0$, while the green
and blue curves correspond to $n=1$ and $n=10$ (effectively $n\rightarrow\infty$),
respectively. The deviations for all the excited even-parity states are almost
identical. Also shown are the crude approximations, $\rho_{0}\approx x/2$
for the ground state, and $\rho_{n}\approx x$ for the (even) excited
states indicated with dotted (black) curves. The latter is fairly
accurate while the former is not accurate beyond extremely minute
values of $\delta$. Also shown are the more refined approximations
given by Eq.~\eqref{eq:Rho0} and Eq.~\eqref{eq:Rhon} for the ground
and excited states, respectively.}
\label{fig:EvenDeviation}
\end{figure}

Let us now investigate the limit of small values of $\delta$.  First consider
the case $\beta\neq0$. For the even-parity solution, after taking
the limit $\delta\rightarrow0$ in Eq.~\eqref{eq:EvenEnergy} and
then simplifying, the result is 
\begin{equation}
2=\frac{1}{\beta}+2\left[2\gamma+\ln\left(\frac{2\delta}{\beta}\right)+\Psi\left(1-\beta\right)\right].\label{eq:RegE}
\end{equation}
Here $\Psi$ denotes the digamma function, which is discussed in Appendix~\ref{sec:App_Digamma}.
Note that, throughout the paper we use $\psi$ to denote a wave function whereas $\Psi$ is the digamma function.
As $\delta\rightarrow0$ the logarithmic term diverges, and thus,
to ensure that the right-hand side of this equation is finite for
any $\delta$, the digamma function must also diverge, and as a result
$\beta\rightarrow n\in\mathbb{N}$. We define the energy scale $E_{0}=-\frac{\hbar^{2}}{2ma_{0}^{2}}$,
which is the same energy scale appearing in the 3D hydrogen atom,
so that the even-parity energy eigenvalues limit to 
\begin{equation}
E\rightarrow-\frac{E_{0}}{n^{2}},\ n\in\mathbb{N},\quad \text{as }\delta\rightarrow0.\label{eq:Energy3}
\end{equation}
In Fig.~\ref{fig:Energy} we plot $\beta\equiv\left(-E_{0}/E\right)^{\frac{1}{2}}$,
as computed from Eq.~\eqref{eq:EvenEnergy} and Eq.~\eqref{eq:OddEnergy}, as a function of $x\equiv1/\ln\left(\frac{1}{\delta}\right)$,
and as can be observed the parameter $\beta$ does indeed converge
to an integer $n$ in the limit $\delta\rightarrow0$. 

To study the deviation of the parameter $\beta$ from an integer $n$ we define
$\rho_{n}=\beta-n$ (Loudon~\citep{Loudon_1959}
calls these ``quantum defects''), which obeys $\rho_{n}\rightarrow0$ as $\delta\rightarrow0$.
The dependence of $\rho_{n}$ on $x$ can be deduced from Eq.~\eqref{eq:RegE}
as follows. First, using the identities in Eqs.~\eqref{eq:DigammaRecRel} and \eqref{eq:DigammaRefForm} in Appendix~\ref{sec:App_Digamma}, we obtain $\Psi\left(1-\beta\right)=\Psi\left(\beta\right)+\pi\cot\left(\pi\beta\right)=\Psi\left(\beta+1\right)-\frac{1}{\beta}+\pi\cot\left(\pi\beta\right).$
Thus, as $\beta\rightarrow n$, we find that $\Psi\left(1-\beta\right)\rightarrow\Psi\left(n+1\right)-\frac{1}{n}+\frac{1}{\rho_{n}}$.
In Eq.~\eqref{eq:DigammaSum}, it is proved that $\Psi\left(n+1\right)=\sum_{k=1}^{n}\frac{1}{k}-\gamma$.
If we define the constant $c_{n}$ by 
\begin{equation}
c_{n}=\gamma-1-\frac{1}{2n}+\ln2+\left(\sum_{k=1}^{n}\frac{1}{k}-\ln n\right),\label{eq:CnCoeff}
\end{equation}
then Eq.~\eqref{eq:RegE} becomes $\frac{1}{\rho_{n}}=\frac{1}{x}\left(1-xc_{n}\right).$
This is a more accurate version of Eq.~(3.26) in Ref.~[\onlinecite{Loudon_1959}],
which, in our notation, is given by $\frac{1}{\rho_{n}}=\frac{1}{x}-\left(\ln2-\ln n\right)$.
Inverting this equation, and then taking the limit $x\rightarrow0$, gives the solution for $\rho_{n}$:
\begin{equation}
\rho_{n}=x\left(1+c_{n}x\right),\ \text{as }x\rightarrow0.\label{eq:Rhon}
\end{equation}
As $n\rightarrow\infty$, the terms in brackets in Eq.~\eqref{eq:CnCoeff}
approach $\gamma$; thus, $c_{n}\rightarrow2\gamma-1+\ln2\approx0.8476$
as $n\rightarrow\infty$. Note that $c_1 \approx 0.7704$ and $c_2 \approx 0.8272$, and this number moves progressively closer to $c_{\infty}$,
so there is not a lot of variation of this constant with $n$.

Now consider the case when $\beta\rightarrow0$. Define $\rho_{0}=\beta.$
In this limit, Eq.~\eqref{eq:RegE} becomes 
\begin{equation}
1=\frac{1}{2\rho_{0}}+\gamma+\ln2-\frac{1}{x}+\ln\left(\frac{1}{\rho_{0}}\right).\label{eq:InvRho0}
\end{equation}
As $\delta\rightarrow0$, i.e., as $x\rightarrow0$, the lowest-order
solution to this equation is $\rho_{0}=\frac{x}{2}$. If we define
the constant $c_{0}$ by 
\begin{equation}
c_{0}=\gamma+2\ln2-1,\label{eq:CoCoeff}
\end{equation}
then after inserting this result into Eq.~\eqref{eq:InvRho0}, along
with replacing the logarithm term by its lowest-order approximation, we
find that
\begin{equation}
\rho_{0}=\frac{x}{2}\left[1+x\left(c_{0}-\ln x\right)\right],\ \text{as }x\rightarrow0.\label{eq:Rho0}
\end{equation}
This is a more accurate version of Eq.~(3.28) in Ref.~[\onlinecite{Loudon_1959}].
As $\delta\rightarrow0$, the energy for the $n=0$ even state becomes
increasingly large and negative. In Fig.~\ref{fig:EvenDeviation} the deviation $\rho_{n}=\beta-n$
is plotted as a function of $x$. It is clear from the figure that Eq.~\eqref{eq:RegE}
provides very accurate results in the range of $\delta$ considered
here, and iterative solution of the much more difficult Eq.~\eqref{eq:EvenEnergy} is not required.

\subsection{Odd-parity solutions}

\begin{figure}[h!]
\centering
\includegraphics[width=10cm]{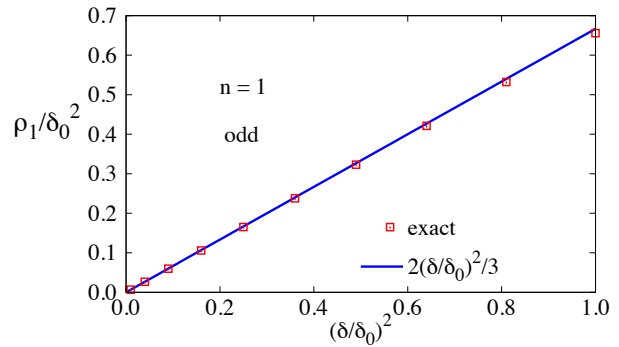}
\caption{The deviation of $\beta$ with respect to integer values, defined
as $\rho_{n}\equiv\beta-n$, versus $\left(\delta/\delta_{0}\right)^{2}$
for the $n=1$ odd state. The exact values are given by the
squares -- these are the same results as shown in Fig.~\ref{fig:Energy}.
The solid curve corresponds to Eq.~\eqref{eq:Rho1Odd}. The parameter $\delta_0=0.01$, 
and $\delta^2$ and $\rho_{1}$ are measured in units of $\delta_{0}^2$.
The values of $\delta$ increment in steps of 0.001, from 0.001 up to 0.01. 
The deviations for the other excited odd states are almost identical.}
\label{fig:OddDeviation}
\end{figure}

First consider the case $\beta\neq0$. For the odd-parity solutions,
the analysis is more involved and requires expanding the $U$ function in Eq.~\eqref{eq:OddEnergy}
to second order (see Eq.~\eqref{eq:Useries} in Appendix~\ref{sec:App_Kummer}), which gives 
\begin{widetext}
\begin{equation}
 -1+\frac{2\delta}{3}-\frac{\delta}{\beta}=
 \frac{\left[\ln\left(\frac{2\delta}{\beta}\right)+\Psi\left(1-\beta\right)-2\Psi\left(1\right)\right]+\left(1-\beta\right)\frac{2\delta}{\beta}\left[\ln\left(\frac{2\delta}{\beta}\right)+\Psi\left(2-\beta\right)-2\Psi\left(2\right)\right]}{\frac{1}{2\delta}-\left[\ln\left(\frac{2\delta}{\beta}\right)+\Psi\left(1-\beta\right)-\Psi\left(1\right)-\Psi\left(2\right)\right]+\left(\beta-1\right)\frac{\delta}{\beta}\left[\ln\left(\frac{2\delta}{\beta}\right)+\Psi\left(2-\beta\right)-\Psi\left(2\right)-\Psi\left(3\right)\right]}.
\label{eq:OddDeltaSeries}
\end{equation}
\end{widetext}
As will be shown, in order to determine the form of $\rho_{n}$ it
is important to retain a portion of all of the terms appearing in the numerator
and denominator. Following the previous section, we let $\beta=\rho_{n}+n$.
Note that $\Psi\left(1-\beta\right)=\Psi\left(\beta\right)+\pi\cot\left(\pi\beta\right)$. 
Thus, as $\rho_{n}\rightarrow0$, for any $n\geq1$, $\Psi\left(1-\beta\right)\rightarrow\Psi\left(n\right)+1/\rho_{n}$.
Similarly, since $\Psi\left(2-\beta\right)=\Psi\left(1-\beta\right)+\frac{1}{1-\beta}$,
then, for any $n\geq2$, $\Psi\left(2-\beta\right)\rightarrow\Psi\left(n-1\right)+1/\rho_{n}$.
Consider the case $n=1$, and on the right-hand side of Eq.~\eqref{eq:OddDeltaSeries} keep only
the singular terms. The result is then 
\begin{equation}
-1+\frac{2\delta}{3}-\delta=\frac{\ln\left(2\delta\right)+\frac{1}{\rho_{1}}}{\frac{1}{2\delta}-\left[\ln\left(2\delta\right)+\frac{1}{\rho_{1}}\right]}+\dots.
\end{equation}
The ellipsis denotes constant terms. Now cross multiply and retain
only the singular terms; to ensure there are no singular terms,
$\rho_{1}$ must be given by 
\begin{equation}
\rho_{1}=\frac{2}{3}\delta^{2}.\label{eq:Rho1Odd}
\end{equation}

Now consider the case $n\geq2$. In this case, the other terms in
the numerator and denominator contribute: 
\begin{equation}
-1+\frac{2\delta}{3}-\frac{\delta}{n}=\frac{\left[\ln\left(\frac{2\delta}{n}\right)+\frac{1}{\rho_{n}}\right]\left[1+\left(1-n\right)\frac{2\delta}{n}\right]}{\frac{1}{2\delta}-\left[\ln\left(\frac{2\delta}{n}\right)+\frac{1}{\rho_{n}}\right]\left[1+\left(1-n\right)\frac{\delta}{n}\right]}.
\end{equation}
Equating both sides requires 
\begin{equation}
\rho_{n}=\frac{2}{3}\delta^{2}.
\end{equation}
Thus, we reach the conclusion that, for all $n\geq1$, $\rho_{n}=\frac{2}{3}\delta^{2}$.
This corrects Eq.~(3.37) in Ref.~[\onlinecite{Loudon_1959}], which did
not include the prefactor of ${1}/{3}$. 

Now consider the case when $\beta\rightarrow0$. In this limit, it is easy to see that Eq.~\eqref{eq:OddDeltaSeries}
has no solution. Thus, there
is no odd-parity state with increasingly negative energy. In Fig.~\ref{fig:OddDeviation} the
deviation $\rho_{1}=\beta-1$ is plotted as a function of $x$. 
The figure shows that Eq.~\eqref{eq:Rho1Odd} provides accurate results in the range of $\delta$ considered
here, and iterative solution of the much more difficult Eq.~\eqref{eq:OddEnergy} is not required. Results for $n > 1$ are essentially identical.

\section{Wave functions}
\label{sec:Wavefunctions}

To determine the constants $A_{\text{I}},B_{\text{I}}$, and $B_{\text{II}}$,
the continuity condition along with the normalization constraint must
be imposed. For the even and odd-parity solutions respectively, continuity of $\psi$ at
$x=\delta a_{0}$ requires 

\begin{align}
A_{\text{I}}\cos\left(qa_{0}\delta\right)=B_{\text{II}}\frac{\delta}{\beta}e^{-\frac{\delta}{\beta}}U\left(1-\beta,2;\frac{2\delta}{\beta}\right),\label{eq:Continuity1}\\
B_{\text{I}}\sin\left(qa_{0}\delta\right)=B_{\text{II}}\frac{\delta}{\beta}e^{-\frac{\delta}{\beta}}U\left(1-\beta,2;\frac{2\delta}{\beta}\right).\label{eq:Continuity2}
\end{align}
Combining Eq.~\eqref{eq:Continuity1} and Eq.~\eqref{eq:Continuity2}
with the normalization condition, $1=\int_{-\infty}^{\infty}\left|\psi\left(x\right)\right|^{2}dx$,
then determines $A_{\text{I}}$ and $B_{\text{I}}$ for the even and odd-parity
solutions respectively. Therefore, the coefficients $A_{\text{I}}$
and $B_{\text{I}}$ are deduced from
\begin{align}
A_{\text{I}}^2 & =  \frac{1}{2a_{0}\delta}\biggl\{\int_{0}^{1}\cos^{2}\left(qa_{0}\delta y\right)dy\nonumber\\
&+\cos^{2}\left(qa_{0}\delta\right)\int_{1}^{\infty}\left[ye^{\frac{\delta}{\beta}\left(1-y\right)}\frac{U\left(1-\beta,2;\frac{2\delta}{\beta}y\right)}{U\left(1-\beta,2;\frac{2\delta}{\beta}\right)}\right]^{2}dy\biggr\}^{-1},\label{eq:Norm1}\\
B_{\text{I}}^2 & =  \frac{1}{2a_{0}\delta}\biggl\{\int_{0}^{1}\sin^{2}\left(qa_{0}\delta y\right)dy\nonumber\\
&+\sin^{2}\left(qa_{0}\delta\right)\int_{1}^{\infty}\left[ye^{\frac{\delta}{\beta}\left(1-y\right)}\frac{U\left(1-\beta,2;\frac{2\delta}{\beta}y\right)}{U\left(1-\beta,2;\frac{2\delta}{\beta}\right)}\right]^{2}dy\biggr\}^{-1}.\label{eq:Norm2}
\end{align}

The even-parity and odd-parity solutions are now completely specified.
In the even-parity case, the eigenvalue $\beta$ is determined from
Eq.~\eqref{eq:EvenEnergy} and the wave function coefficients are
obtained from Eqs.~\eqref{eq:Continuity1}  and \eqref{eq:Norm1}; in the
odd-parity case, the eigenvalue $\beta$ is determined from Eq.~\eqref{eq:OddEnergy}
and the coefficients of the wave function are obtained from Eqs.~\eqref{eq:Continuity2} and \eqref{eq:Norm2}. 

In the limit that $\delta\rightarrow0$, the expected form of the wave functions
can be deduced as follows. Consider first the case of the odd-parity
eigenstates. In Eq.~(13.6.27) of Ref.~[\onlinecite{AS_Book}], the following result
is given: $U\left(-n,1+\alpha;x\right)=\left(-1\right)^{n}n!L_{n}^{\left(\alpha\right)}\left(x\right)$.
Thus, the (normalized) odd-parity eigenstates limit to 
\begin{equation}
\psi_{\text{odd}}\left(x\right)\rightarrow\left(-1\right)^{n-1}\left(\frac{2}{a_{0}n^{3}}\right)^{\frac{1}{2}}\frac{x}{a_{0}n}e^{-\left|x\right|/\left(a_{0}n\right)}L_{n-1}^{\left(1\right)}\left(\frac{2\left|x\right|}{a_{0}n}\right),\text{\ }n\geq1.\label{eq:Odd_Psi}
\end{equation}
The proof of the normalization constant is provided in Appendix~\ref{sec:App_Laguerre}.
This result agrees with Eq.~(A6) in Ref.~[\onlinecite{Palma_2006}] and
with Eq.~(3.29) in Ref.~[\onlinecite{Loudon_1959}] (noting the difference in
definitions of the associated Laguerre polynomials; see Appendix~\ref{sec:App_Laguerre}). 

\begin{figure}
\centering
\includegraphics[width=10cm]{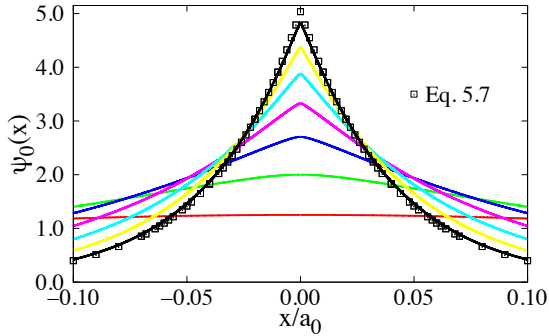}
\caption{The even-parity ground state for $\delta=10^{-1}$ (red), $10^{-2}$ (green), $10^{-3}$ (blue), $10^{-4}$ (mauve), $10^{-5}$ (cyan), $10^{-6}$ (yellow), 
and $10^{-7}$ (black dash-dot).
The analytical result, denoted by the squares for the smallest value of $\delta$ only, 
is given by Eq.~\eqref{eq:Even_Psi0}. Note the scale of the horizontal scale --- we are zooming into domain near
the origin.
The ground state clearly varies significantly with $\delta$. For
the smallest value of $\delta$, the numerical result is slightly
different near the origin from the analytical result. This is to be
expected, since the former has a continuous derivative at $x=0$ while
the latter does not.}
\label{fig:n0Wavefunction}
\end{figure}

Now consider the even-parity eigenstates. As $\delta\rightarrow0$,
for $n\geq1$, the even-parity eigenstates limit to $\psi_{\text{even}}\left(x\right)\rightarrow\psi_{\text{odd}}\left(\left|x\right|\right)$.
For the $n=0$ even-parity eigenstate, we use Eq.~\eqref{eq:Useries} in Appendix~\ref{sec:App_Kummer}:
$U\left(1,2;z\right)=1/z$. Therefore, the normalized even-parity eigenstates have the limiting form

\begin{eqnarray}
\psi_{\text{even}}\left(x\right) & \rightarrow & \psi_{\text{odd}}\left(\left|x\right|\right),\text{\ }n\geq1.\label{eq:Even_Psi}\\
\psi_{0}\left(x\right) & \rightarrow & \lim_{\beta\rightarrow0}\frac{1}{\sqrt{a_{0}\beta}}e^{-\left|x\right|/\left(a_{0}\beta\right)},\ n=0.\label{eq:Even_Psi0}
\end{eqnarray}
The wave function in Eq.~\eqref{eq:Even_Psi0} was shown in Refs.~[\onlinecite{Andrews_1966,Yepez_1987,Xianxi_1997}]
to have zero overlap with any square-integrable function
$\phi\in L_{2}\left(\mathbb{R}\right)$. On the other hand, in Ref.~[\onlinecite{Loudon_1959}] 
the square of $\psi_{0}$ was identified as limiting towards a
Dirac-delta function $\psi_{0}^{2}\rightarrow\delta\left(x\right)$; see Ref.~[\onlinecite{Yepez_1987}] for a discussion of this dichotomy.
In Fig.~\ref{fig:n0Wavefunction} we show the ground-state wave function for various values of the regularization parameter $\delta$. The
ground-state wave function is highly peaked at the origin, and Fig.~\ref{fig:n0Wavefunction} provides a zoomed-in view. For the smallest
value of $\delta$ we have also plotted Eq.~\eqref{eq:Even_Psi0} (squares), and the agreement with our result is very good. Nonetheless we should point
out that Eq.~\eqref{eq:Even_Psi0} has a cusp at the origin, whereas our regularized solutions (curves in Fig.~\ref{fig:n0Wavefunction}) do not.
In fact, according to Ref.~[\onlinecite{Palma_2006}], the non-regularized 1D Coulomb
potential, which has $\delta=0$ in our notation, has no even-parity
eigenstates, since such solutions would have a discontinuous first derivative. 

\begin{figure}
\centering
\includegraphics[width=10cm]{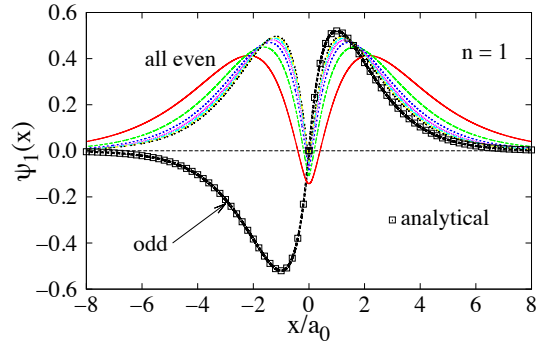}
\caption{The $n=1$ even and odd-parity energy eigenstates 
for $\delta=10^{-1}$ (red), $10^{-2}$ (green), $10^{-3}$ (blue), $10^{-4}$ (mauve), $10^{-5}$ (cyan), $10^{-6}$ (yellow), and $10^{-7}$ (black dash-dot).
The analytical results, denoted by the squares, are given by Eq.~\eqref{eq:Odd_Psi}
and Eq.~\eqref{eq:Even_Psi}, and we have shown only the $x>0$ half of Eq.~\eqref{eq:Even_Psi} for clarity. For all values of $\delta$, the
odd-parity eigenstates cannot be distinguished from one another (all 7 different curves are shown in black and coincide with the points)
The even-parity eigenstates,
however, vary more significantly with $\delta$, and even for the smallest
value of $\delta$ the numerical result is still slightly different from
the analytical result, as is clear from the plot for $x > 0$. Nonetheless, as $\delta$ continues to decrease, the numerical results
will coincide with the analytical result indicated by Eq.~\eqref{eq:Even_Psi}, and shown in the figure for $x>0$.}
\label{fig:n1Wavefunction}
\end{figure}

\begin{figure}
\centering
\includegraphics[width=10cm]{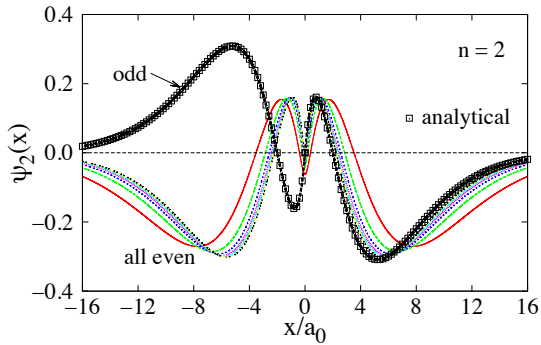}
\caption{The $n=2$ even and odd-parity energy eigenstates for $\delta=10^{-1}$ (red), $10^{-2}$ (green), $10^{-3}$ (blue), $10^{-4}$ (mauve), $10^{-5}$ (cyan), $10^{-6}$ (yellow), and $10^{-7}$ (black dash-dot).
The analytical result for the odd state, denoted by the squares, is given by Eq.~\eqref{eq:Odd_Psi}, and we have shown only the $x>0$ half of Eq.~\eqref{eq:Even_Psi} for clarity. As was the case with $n=1$, for all values of $\delta$, the
odd-parity eigenstates cannot be distinguished from one another all 7 different curves are shown in black and coincide with the points.) The even-parity eigenstates,
however, vary more significantly with $\delta$, and even for the smallest
value of $\delta$ the numerical result is still slightly different from
the analytical result, as is clear from the plot for $x > 0$. Nonetheless, as $\delta$ continues to decrease, the numerical results
will coincide with the analytical result indicated by Eq.~\eqref{eq:Even_Psi}, and shown in the figure for $x>0$.}
\label{fig:n2Wavefunction}
\end{figure}

This observation is also hinted at in Figs.~\ref{fig:n1Wavefunction} and \ref{fig:n2Wavefunction} where we
plot the $n=1$ and $n=2$ wave functions for the regularized 1D Coulomb potential along
with the limiting analytical form given in Eq.~\eqref{eq:Odd_Psi}. We have plotted the analytical form of only the odd solution, and as Eq.~\eqref{eq:Even_Psi}
indicates, this coincides with the even analytical solution for $x>0$. The even analytical solution for $x<0$ is the mirror image of this solution (not shown), so that here
too a cusp is present at the origin for the analytical even excited states. 
While the even results for non-zero regularization parameter $\delta$ trend towards the analytical result, they {\it also} appear to have a cusp-like feature near $x=0$. Of course they do not, as the small $x$ behavior is given by the cosine function. This plot illustrates, in a practical way, how finite-$\delta$ results can nonetheless begin to reproduce features of the $\delta=0$ result. In the meantime, ignoring for the moment what is occurring very close to the origin, the rest of the even wave function is slowly approaching the expected analytical behavior given by Eq.~\eqref{eq:Even_Psi}.

The situation with the odd-parity states is very different. Over the 6 decades of variation of the regularization parameter $\delta$, the wave function has already
converged to the $\delta = 0$ solution given by Eq.~\eqref{eq:Odd_Psi}; indeed, the different curves are not discernible on this scale.
These subtleties can be explained as follows.
For the regularized 1D Coulomb potential, the energy eigenstates belong
to the subspace of square-integrable functions with continuous first derivative.
For the singular 1D coulomb potential, however, the Hamiltonian is self-adjoint only on the space of functions that vanish at the origin and have a continuous first derivative.~\citep{Kurasov_1996,Palma_2006}
Our results for the regularized problem show that, while for any $\delta>0$ the even-parity eigenstates are indeed $C^{1}$, they limit to a function that is not $C^{1}$ at $x=0$. 
The odd-parity eigenstates, however, limit to a $C^{1}$ function.  
In other words, we have the peculiar circumstance that the two models show a discontinuous behavior for the even-parity states as
the regularization parameter $\delta \rightarrow 0$. While both are mathematically tractable, we believe the regularized model is physically appropriate.

As a result, there are two fundamentally different systems: a regularized 1D Coulomb potential, with even and odd-parity solutions, and a singular Coulomb potential. 
That these two systems have different eigenspectra is an interesting observation, one which is the root of the cause for the disputes in the literature concerning the validity of the even-parity solutions for the singular 1D Coulomb potential. 
Nonetheless, what is clear from our progression of results with decreasing $\delta$ is that all eigenvalues and eigenstates remain
well-defined and ``proper,'' no matter how small $\delta$ is taken. Moreover, they all numerically approach the limits inferred from the mathematical
limit $\delta \rightarrow 0$. A more nuanced discussion is required for the ground state, since there is really no analytical result, only limiting behavior as indicated by
Eq.~\eqref{eq:Rho0} for the eigenvalue and Eq.~\eqref{eq:Even_Psi0} for the eigenstate. Since both the eigenvalue and the eigenstate are singular
in this limit, it is not surprising that these are inaccessible from the bare $\delta =0$ hydrogen potential.\\

\section{Conclusions}
\label{sec:Conclusions}

In this paper we have analyzed a regularized version of the one-dimensional
hydrogen atom consisting of a potential that is constant in the vicinity
of the origin, and Coulomb-like beyond. We have obtained results very much in agreement with those obtained by Loudon \citep{Loudon_1959} 
for a different regularization. This system has both even and odd-parity eigenstates,
and moreover, for any finite cutoff, the eigenstates are nondegenerate. Nonetheless, as our regularization
parameter representing the cutoff near the origin, $\delta$, approaches zero, the even-parity eigenvalues approach those
of their odd-parity counterparts. Their wave functions remain well-defined, and also approach their parity-adjusted odd-parity counterparts.
The most intriguing feature of this model is the even-parity ground state, whose energy becomes increasingly negative as $\delta \rightarrow 0$.
Concomitant with this behavior, the corresponding ground state becomes more localized near the origin, approaching a functional form whose
square approaches that of a Dirac $\delta$-function.
Because of the so-called pseudo-potential effect~\citep{Ibrahim_2018} the presence of this
state gives rise to an effective barrier for all other states, a property recognized by Andrews in Ref.~[\onlinecite{Andrews_1976}]. 
This barrier serves to organize the remaining eigenstates into split even and odd doublets, as expected for a simple double well. The strength
of this effective double well barrier is controlled by the regularization parameter $\delta$.

\begin{acknowledgments}
This work was supported in part by the Natural Sciences and Engineering Research Council of Canada (NSERC). R.B. acknowledges support from the Department of Physics and the Theoretical Physics Institute at the University of Alberta.
\end{acknowledgments}

\appendix
\numberwithin{equation}{section}
\section{Gamma and Digamma functions}
\label{sec:App_Digamma}

In the following appendices we provide a brief overview of the special
functions and their pertinent identities used in the manuscript. The
Gamma function was defined by Weierstrass (see Ch.~12 of Ref.~[\onlinecite{Whittaker_WatsonBook}] for example) by the equation 
\begin{equation}
\frac{1}{\Gamma\left(z\right)}=ze^{\gamma z}\prod_{n=1}^{\infty}\left[\left(1+\frac{z}{n}\right)e^{-\frac{z}{n}}\right].\label{eq:WeierstrassExp}
\end{equation}
Here, $z$ is a complex number not equal to zero or a negative integer, and $\gamma$
is the Euler-Mascheroni constant, which approximates to $\gamma\approx0.5772$,
and is defined exactly by 
\begin{equation}
\gamma=\lim_{n\rightarrow\infty}\left(\sum_{k=1}^{n}\frac{1}{k}-\ln\left(n\right)\right).
\end{equation}
Using Eq.~\eqref{eq:WeierstrassExp}, along with its derivative with
respect to $z$, after setting $z=1$ we obtain 
\begin{eqnarray}
\Gamma\left(1\right) & = & 1.\\
\Gamma^{\prime}\left(1\right) & = & -\lim_{n\rightarrow\infty}\sum_{k=1}^{n}\left(\frac{1}{k}-\ln\left(1+\frac{1}{k}\right)\right) = -\gamma.
\end{eqnarray}
An integral representation of the Gamma function, due to Euler, is 
\begin{equation}
\Gamma\left(z\right)=\int_{0}^{\infty}dtt^{z-1}e^{-t}.\label{eq:EulerExp}
\end{equation}
Two important properties of the Gamma function are given by
\begin{eqnarray}
\Gamma\left(z+1\right) & = & z\Gamma\left(z\right).\label{eq:GammaRecRel}\\
\Gamma\left(1-z\right)\Gamma\left(z\right) & = & \frac{\pi}{\sin\left(\pi z\right)}.\label{eq:GammaRefForm}
\end{eqnarray}
The first identity is easily proved using Eq.~\eqref{eq:EulerExp}.
The second identity, which is known as the reflection formula, can
be proved using Eq.~\eqref{eq:WeierstrassExp} and Eq.~\eqref{eq:GammaRecRel} and the Weierstrauss factorization formula for the sine function. In the case where $z\in\mathbb{N}$,
Eq.~\eqref{eq:GammaRecRel} becomes $\Gamma\left(n+1\right)=n!$.
Another useful identity is known as the duplication formula, given by 
\begin{equation}
\pi^{\frac{1}{2}}\Gamma\left(2z\right)=2^{2z-1}\Gamma\left(z\right)\Gamma\left(z+\frac{1}{2}\right).
\end{equation}
 
The digamma function is defined by 
\begin{equation}
\Psi\left(z\right)=\frac{d}{dz}\ln\Gamma\left(z\right).
\end{equation}
By taking the logarithmic derivatives of Eq.~\eqref{eq:GammaRecRel}
and Eq.~\eqref{eq:GammaRefForm}, two recurrence relations for the
digamma function are obtained:
\begin{eqnarray}
\Psi\left(z+1\right) & = & \Psi\left(z\right)+\frac{1}{z}.\label{eq:DigammaRecRel}\\
\Psi\left(1-z\right)-\Psi\left(z\right) & = & \pi\cot\left(\pi z\right).\label{eq:DigammaRefForm}
\end{eqnarray}
The first identity above proves useful in evaluating the partial sums of the Harmonic series:
\begin{eqnarray}
\sum_{k=1}^{n}\frac{1}{k} & = & \sum_{k=1}^{n}\left[\Psi\left(k+1\right)-\Psi\left(k\right)\right]\nonumber \\
 & = & \Psi\left(n+1\right)-\Psi\left(1\right)\nonumber \\
 & = & \Psi\left(n+1\right)+\gamma.\label{eq:DigammaSum}
\end{eqnarray}
In addition, the limiting behavior of $\Psi\left(z\right)$, as $z\rightarrow-n$, where $n\in\mathbb{N}$, can be deduced from Eq.~\eqref{eq:DigammaRefForm} as 
\begin{equation}
\Psi\left(z\right)\rightarrow-\frac{1}{z+n}.
\end{equation}
The asymptotic behavior of the digamma function is (see Eq.~(6.3.18) in Ref.~[\onlinecite{AS_Book}])
\begin{equation}
\Psi\left(z\right)\rightarrow\ln\left(z\right)-\frac{1}{2z}-\frac{1}{12z^{2}}+O\left(z^{-4}\right).
\end{equation}

\section{Hypergeometric and Tricomi functions}
\label{sec:App_Kummer}

The confluent hypergeometric differential equation, also known as
Kummer's equation, for the function $f\left(z\right)$ is given by (see Ch.~13 of Ref.~[\onlinecite{AS_Book}] and Ch.~13 of Ref.~[\onlinecite{Olver_Book}]):
\begin{equation}
z\frac{d^{2}f}{dz^{2}}+\left(b-z\right)\frac{df}{dz}-af=0.
\end{equation}
The differential equation has a regular
singularity at $z=0$ and an irregular singularity at $z=\infty$.
The two linearly independent solutions of interest are known as Kummer's
function $M\left(a,b,z\right)$ and Tricomi's function $U\left(a,b,z\right)$.
The $M\left(a,b,z\right)$ power series representation about $z=0$
is given by 
\begin{equation}
M\left(a,b,z\right)=1+\frac{a}{b}z+\frac{\left(a\right)_{2}}{\left(b\right)_{2}}\frac{z^{2}}{2!}+\cdots+\frac{\left(a\right)_{n}}{\left(b\right)_{n}}\frac{z^{n}}{n!}+\dots,
\end{equation}
where $\left(a\right)_{n}=a\left(a+1\right)\left(a+2\right)\dots\left(a+n-1\right)$,
$\left(a\right)_{0}=1$. Similarly, for $U\left(a,b,z\right)$ we have 
\begin{align}
U\left(a,b,z\right)&=\frac{\pi}{\sin\left(\pi b\right)}\biggl[\frac{M\left(a,b,z\right)}{\Gamma\left(1+a-b\right)\Gamma\left(b\right)}\nonumber\\
&\quad-z^{1-b}\frac{M\left(1+a-b,2-b,z\right)}{\Gamma\left(a\right)\Gamma\left(2-b\right)}\biggr].
\end{align}
The Tricomi function is a many-valued function, and its principal
branch is given by $-\pi<\text{arg}\leq\pi$. The logarithmic series for $U$ is given in Eq.~(13.2.9) of Ref.~[\onlinecite{Olver_Book}]:
\begin{align}
U\left(a,n+1,z\right) & =  \frac{\left(-1\right)^{n+1}}{n!\Gamma\left(a-n\right)}
\sum_{k=0}^{\infty}\frac{\left(a\right)_{k}z^{k}}{\left(n+1\right)_{k}k!}\biggl[ \ln z \nonumber\\
 & + \Psi\left(a+k\right)-\Psi\left(1+k\right)-\Psi\left(n+k+1\right)\biggr] \nonumber \\
 & + \frac{1}{\Gamma\left(a\right)}\sum_{k=1}^{n}\frac{\left(k-1\right)!\left(1-a+k\right)_{n-k}}{\left(n-k\right)!}z^{-k}.
 \label{eq:Useries}
\end{align}
The most pertinent recurrence relations for the $M$ and $U$ functions
that were used in the main text are listed below; a more detailed
collection of identities can be found in Refs.~[\onlinecite{AS_Book}] and [\onlinecite{Olver_Book}]. From Eq.~(13.4.21) in Ref.~[\onlinecite{AS_Book}], we have 
\begin{equation}
\frac{d}{dz}U\left(a,b,z\right)=-aU\left(a+1,b+1,z\right).
\end{equation}
In addition, Eqs.~(13.4.17-18) in Ref.~[\onlinecite{AS_Book}] are given by 
\begin{align}
U\left(a,b,z\right)-aU\left(a+1,b,z\right)-U\left(a,b-1,z\right) & =  0,\label{eq:Urel1}\\
\left(b-a\right)U\left(a,b,z\right)+U\left(a-1,b,z\right)-zU\left(a,b+1,z\right) & =  0\label{eq:Urel2}.
\end{align}
The asymptotic behavior of the $M$ and $U$ functions are written
below. As $\left|z\right|\rightarrow\infty$, 
\begin{align}
M\left(a,b,z\right) & \rightarrow  \frac{\Gamma\left(b\right)}{\Gamma\left(a\right)}e^{z}z^{a-b}\left[1+O\left(\left|z\right|^{-1}\right)\right],\ \text{Re}z>0,\\
M\left(a,b,z\right) & \rightarrow  \frac{\Gamma\left(b\right)\left(-z\right)^{-a}}{\Gamma\left(b-a\right)}\left[1+O\left(\left|z\right|^{-1}\right)\right],\ \text{Re}z<0.
\end{align}
These expressions assume that $a$ is not a negative integer, in that
case the $M$ function truncates to a polynomial; this is discussed
further in the next section. As $\text{Re}z\rightarrow\infty$, 
\begin{equation}
U\left(a,b,z\right)\rightarrow z^{-a}\left[1+O\left(\left|z\right|^{-1}\right)\right].
\end{equation}

\section{Laguerre polynomials}
\label{sec:App_Laguerre}

Following Eq.~(22.11.6) in Ref.~[\onlinecite{AS_Book}], we define the associated Laguerre polynomials according to Rodrigues' formula:
\begin{equation}
L^{\left(\alpha\right)}_{n}\left(x\right)=\frac{1}{n!}e^xx^{-\alpha}\frac{d^n}{dx^n}\left(e^{-x}x^{n+\alpha}\right).
\end{equation}
This definition agrees with that used in Eq.~(A5) of Ref.~[\onlinecite{Palma_2006}] (although that reference uses a calligraphic $\mathcal{L}$ whereas we use the italicized latin $L$).
Note, however, this definition differs from the associated Laguerre polynomials defined in Ref.~[\onlinecite{Griffiths_Book}].
In certain special cases the $M$ and $U$ functions reduce to polynomial
solutions. For the purposes of this paper, the pertinent case is when
$M$ and $U$ reduce to the associated Laguerre polynomials (see Eqs.~(13.6.9) and (13.6.27), respectively, of Ref.~[\onlinecite{AS_Book}]): 
\begin{eqnarray}
M\left(-n,\alpha+1,x\right) & = & \frac{n!}{\left(\alpha+1\right)_{n}}L_{n}^{\left(\alpha\right)}\left(x\right),\\
U\left(-n,\alpha+1,x\right) & = & \left(-1\right)^{n}n!L_{n}^{\left(\alpha\right)}\left(x\right).
\end{eqnarray}

As an example of the utility of these various identities, we prove that the wave function in Eq.~(\ref{eq:Odd_Psi}) is normalized. The wave function is 
\begin{equation}
\psi_{n}\left(x\right)=\left(-1\right)^{n-1}\left(\frac{2}{a_{0}n^{3}}\right)^{\frac{1}{2}}\frac{\left|x\right|}{a_{0}n}e^{-\left|x\right|/\left(a_{0}n\right)}L_{n-1}^{\left(1\right)}\left(\frac{2\left|x\right|}{a_{0}n}\right),\text{\ }n\geq1.
\end{equation}
The integral of the probability density is then 
\begin{eqnarray}
N&=&\int_{-\infty}^{\infty}\left|\psi_{n}\left(x\right)\right|^{2}dx \nonumber\\
& = & \frac{4}{a_{0}n^{3}}\int_{0}^{\infty}\left(\frac{x}{a_{0}n}\right)^{2}e^{-2x/\left(a_{0}n\right)}\left[L_{n-1}^{\left(1\right)}\left(\frac{2\left|x\right|}{a_{0}n}\right)\right]^{2}dx\nonumber \\
 & = & \frac{1}{2n^{2}}\int_{0}^{\infty}e^{-u}\left[uL_{n-1}^{\left(1\right)}\left(u\right)\right]^{2}du.\label{eq:NormInt1}
\end{eqnarray}
To evaluate this integral, we need the following two identities, which
correspond to Eq.~(22.7.30) and Eq.~(22.8.6), respectively, of Ref.~[\onlinecite{AS_Book}]:
\begin{eqnarray}
L_{n}^{\left(\alpha-1\right)}\left(x\right) & = & L_{n}^{\left(\alpha\right)}\left(x\right)-L_{n-1}^{\left(\alpha\right)}\left(x\right).\label{eq:LaguerreId1}\\
x\frac{d}{dx}L_{n}^{\left(\alpha\right)}\left(x\right) & = & nL_{n}^{\left(\alpha\right)}\left(x\right)-\left(n+\alpha\right)L_{n-1}^{\left(\alpha\right)}\left(x\right).
\end{eqnarray}
Combining these two results, we obtain $\frac{d}{du}uL_{n-1}^{\left(1\right)}\left(u\right)=nL_{n-1}^{\left(0\right)}\left(u\right)$.
Using this identity, along with integration by parts, Eq.~\eqref{eq:NormInt1}
then simplifies to
\begin{eqnarray}
N & = & \frac{1}{2n^{2}}\int_{0}^{\infty}e^{-u}\frac{d}{du}\left[uL_{n-1}^{\left(1\right)}\left(u\right)\right]^{2}du\nonumber \\
 & = & \frac{1}{n}\int_{0}^{\infty}e^{-u}uL_{n-1}^{\left(1\right)}\left(u\right)L_{n-1}^{\left(0\right)}\left(u\right)du.\label{eq:NormInt2}
\end{eqnarray}
To evaluate this integral, another recurrence relation is needed (see Eq.~(22.7.32) in Ref.~[\onlinecite{AS_Book}]):
\begin{equation}
L_{n}^{\left(\alpha-1\right)}\left(x\right)=\frac{1}{n+\alpha}\left[\left(n+1\right)L_{n+1}^{\left(\alpha\right)}\left(x\right)-\left(n+1-x\right)L_{n}^{\left(\alpha\right)}\left(x\right)\right].
\end{equation}
Rewriting this equation with $\alpha = 1$ and $n \rightarrow n-1$, and then isolating the last term containing $xL_{n-1}^{(1)}(x)$, results in
\begin{eqnarray}
xL_{n-1}^{(1)}(x) &= &nL_{n-1}^{(0)}(x) - n\left( L_{n}^{(1)}(x) - L_{n-1}^{(1)}(x))\right), \nonumber \\
&= &n\left(L_{n-1}^{(0)}(x) - L_n^{(0)}(x)\right).
\end{eqnarray}
The second line follows from using Eq.~\eqref{eq:LaguerreId1}.
Substituting this identity into (the second line of) Eq.~\eqref{eq:NormInt2} we obtain
\begin{equation}
N  =  \int_{0}^{\infty}e^{-u}L_{n-1}^{(0)}(u) \left[L_{n-1}^{(0)}(u) - L_{n}^{(0)}(u)\right]du = 1.
\end{equation}
The last equality follows from the orthonormality condition of the associated
Laguerre polynomials: $\int_{0}^{\infty}dxx^{\alpha}e^{-x}L_{n}^{\left(\alpha\right)}\left(x\right)L_{m}^{\left(\alpha\right)}\left(x\right)=\frac{1}{n!}\delta_{nm}\Gamma\left(n+\alpha+1\right).$

\end{document}